\documentclass[12pt]{article}
\usepackage{amsfonts}
\usepackage{mitpress}

\newdimen\dummy
\dummy=\oddsidemargin
\input{tcilatex}

\begin{document}

\begin{center}
\bigskip 

{\Huge An axiomatic theory for interaction between species in ecology:
Gause's exclusion conjecture}
\end{center}

\[
\]

\begin{center}
{\Large \ J. C. Flores}$^{a,b}$

$^{a}$Universidad de Tarapac\'{a}, Intituto de Alta Investigaci\'{o}n,
Casilla 7-D, Arica-Chile

$^{b}$CIHDE, Centro de Investigaciones del Hombre en el Desierto, Casilla
7-D, Arica-Chile
\end{center}

\[
\]

\textbf{Abstract: } I introduce an axiomatic representation, called ecoset,
to consider interactions between species in ecological systems. For
interspecific competition, the exclusion conjecture (Gause) is put in a
symbolic way and used as a basic operational tool to consider more complex
cases like two species with two superposed resources (niche differentiation)
and apparent competitors. Competition between tortoises and invaders in
Galapagos islands is considered as a specific example. Our theory gives us
an operative language to consider elementary process in ecology and open a
route to more complex systems.

\[
\]

{\Large I Introduction}

\[
\]

After Ryall (2006), theoretical works in ecology are almost completely
ignored for empirical searchers. The cause becomes related to
inaccessibility of theoretical studies because its presentation. In fact,
most of the time, a theoretical paper requires a specialized background
knowledge. In this sense, a simple technical language in ecology could be
desirable. The purpose of this paper is related with an axiomatic
nomenclature (set theory) to describe ecological process and consider, also,
predictive events. More important, inference rules are proposed,
particularly, the exclusion conjecture is put in an axiomatic way.

Gause's Exclusion Conjecture GEC (or competitive exclusion conjecture)
widely cited in ecology states: \textit{when we have two different species }$%
A$\textit{\ and }$B$\textit{\ competing for the same invariable (temporal
and spatial) ecological primary resource }$\mathcal{S}$\textit{\ (niche),
then one of them disappear} (Begon \textit{et al} (1996), Emmel (1973),
Gause (1934), Hasting (1997)). It is important to note that GEC (which
refers to species, not to process) holds when no migration, no mutation and
no resource (niche) differentiation exist in the invariable ecological
systems. Moreover, it refers to interspecific competition. Applications
could be found in many texts of ecology (see for instance Odum (1971)). A
direct application of this conjecture to Neanderthals extinction in Europe
could be found in references Flores (1998) and Murray (2002). New studies
about Neanderthals and Modern Human competition could be found in Mellars
(2006).

\[
\]

{\Large II Symbology for depredation and competition: basic definitions}

\[
\]

Consider a primary resource $\mathcal{S}$ and species $A,B,C...\phi $
(absence of species). The notation

\begin{equation}
A>B,\text{ \ \ \ (}B\text{ \ consumes }A\text{),}
\end{equation}
means, species $B$ exploits species $A$ as a source in a sense of
depredation. For two no-interacting species $A$ and $B$ (also for ecological
process) in a given region we write $A\oplus B$. So, when two species ($A$
and $B$) use the same primary resource ($\mathcal{S}$), in a not depending
way, we write

\begin{equation}
\mathcal{S}>\left( A\oplus B\right) ,\text{ \ \ \ (independent depredation).}
\end{equation}

The symbol $\Leftrightarrow $ will be interpreted as equivalence between
ecological process or species. In the process (2) we always \ assume
implicitly the equivalence: 
\begin{equation}
\left\{ \mathcal{S}>\left( A\oplus B\right) \right\} \Leftrightarrow \left\{
\left( \mathcal{S}>A\right) \oplus \left( \mathcal{S}>B\right) \right\} .
\end{equation}
An ecological process like $\mathcal{S}>A>B$ does not mean $\mathcal{S}>B$
(no transitivity). If also $B$ consumes $\mathcal{S}$ we must write: $\left( 
\mathcal{S}>A>B\right) \oplus \left( \mathcal{S}>B\right) $.

To consider species in competition (no depredation) we will use the symbol $%
\supset \subset $ (see later). Here we give some basic definitions, for
instance, consider species $A$ and $B$ in struggle for some source like
water, space, refuge, etc. The symbol, 
\begin{equation}
A\supset B,\text{ \ \ \ (}A\text{ perturbed by }B\text{),}
\end{equation}
means that species $B$ perturbs (interferes) $A$. We note that for
depredation we use other symbol ($>$). With these basic definitions we can
represent competition between species. In fact, if $A$ and $B$ are two
species in competition (no depredation) then we write

\begin{equation}
\left( A\supset B\right) \oplus \left( B\supset A\right) ,\text{ \ \ \ (}A%
\text{ and }B\text{ compete).}
\end{equation}

For simplicity we will use the alternative symbol $\supset \subset $ for
competition. Namely, 
\begin{equation}
\left\{ \left( A\supset B\right) \oplus \left( B\supset A\right) \right\}
\Leftrightarrow \left( A\supset \subset B\right) ,\text{ \ \ \ (symbol for
competition).}
\end{equation}

Before to ending this section, we give a useful definition. The notion of 
\textit{potential competitors} is related to two species which put together
then ($\Rightarrow $) compete. In our symbolic notation: 
\begin{equation}
\left\{ \mathcal{S}>\left( A\oplus B\right) \right\} \Rightarrow \left\{ 
\mathcal{S}>\left( A\supset \subset B\right) \right\} ,\text{\ \ \
(potential competitors).}
\end{equation}
Where the symbol $\Rightarrow $ has a temporal interpretation or it defines
a temporal direction.

\[
\]

{\Large III Gause's Exclusion Conjecture (GEC) }

\[
\]
In our notation the Gause`s exclusion conjecture, discussed in section I, is
written as the inference rule,

\begin{equation}
\left\{ \mathcal{S}>\left( A\supset \subset B\right) \right\} \text{ }%
\Rightarrow \left\{ \left( \mathcal{S}>A\right) \text{ \ or }\left( \mathcal{%
S}>B\right) \right\} ,\text{ \ (GEC).}
\end{equation}
The above statement will be a basic operational tool to consider more
general cases or applications like two sources and two predators or,
eventually, others (Flores (2005)). The more famous example of exclusion
comes from the classic laboratory work of Gause (1934), who considers two
type of \textit{Paramecium}, namely, \textit{P. caudatum} and \textit{P.
aurelia}. Both species grow well alone and reaching a stable carrying
capacities in tubes of liquid medium and consuming bacteria and oxygen. When
both species grow together, \textit{P. caudatum} declines to the point of
extinction and leaving \textit{P. aurelia} in the niche.

As said before, other examples of GEC could be found in literature. For
instance, competition between \textit{Tribolium confusum }and \textit{%
Tribolium castaneum} where one species is always eliminated when put
together (Park (1954)). In this case is the temperature and humidity \ which
determine usually the winner. In fact, in a medium range of these parameters
is only a probabilistic approach which determines the victorious species.

To ending this section, a caution note, in the case of \ \textquotedblleft
competition between process\textquotedblright\ usually GEC does not hold
directly.

\[
\]

{\Large IV \ Two species and two resources}

\[
\]

Consider two resources $\mathcal{S}_{1}$ \ and $\mathcal{S}_{2}$ and two
species $A $ and $B$ in competition: $\left( \mathcal{S}_{1}\oplus \mathcal{S%
}_{2}\right) >\left( A\supset \subset B\right) $, or $\left\{ \mathcal{S}%
_{1}>\left( A\supset \subset B\right) \right\} \oplus \left\{ \mathcal{S}%
_{2}>\left( A\supset \subset B\right) \right\} $. Using directly GEC (8) we
have the ecological process:

\[
\left( \mathcal{S}_{1}\oplus \mathcal{S}_{2}\right) >\left( A\supset \subset
B\right) \Rightarrow 
\]

\begin{equation}
\Rightarrow \left\{ \left( \mathcal{S}_{1}>A\right) \text{ or\ }\left( 
\mathcal{S}_{1}>B\right) \right\} \oplus \left\{ \left( \mathcal{S}%
_{2}>A\right) \text{ or }\left( \mathcal{S}_{2}>B\right) \right\} ,
\end{equation}
and then the excluding final states:

(a) $\left\{ \mathcal{S}_{1}\oplus \mathcal{S}_{2}\right\} >A.$ Species\ $A$
exterminates $B$.

(b) $\left\{ \mathcal{S}_{1}\oplus \mathcal{S}_{2}\right\} >B.$ Species $B$
exterminates $A$.

(c) $\left\{ \mathcal{S}_{1}>A\right\} \oplus \left\{ \mathcal{S}%
_{2}>B\right\} $. Species $A$ exploits $\mathcal{S}_{1}$\ and $B$ exploits $%
\mathcal{S}_{2}$.

(d) $\left\{ \mathcal{S}_{1}>B\right\} \oplus \left\{ \mathcal{S}%
_{2}>A\right\} $. Species $A$ exploits $\mathcal{S}_{2}$\ and $B$ exploits $%
\mathcal{S}_{1}$. \ \ \ \ \ \ \ \ \ \ \ \ \ \ \ \ \ \ \ \ \ \ \ \ \ \ \ \ \
\ \ \ \ \ \ \ \ \ \ \ \ \ \ \ \ \ \ \ \ \ \ \ \ \ \ \ 

\ The last two possibilities (c) and (d) \ tell us that both species could
survives by resource exploitations in a differential (partitioned) way.

The results found in the above process (a-d), have been observed in
laboratories where two diatom species (\textit{Asterionella formosa }and%
\textit{\ Cyclotella meneghimiana}) compete for silicate ($\mathcal{S}_{1}$)
and phosphate ($\mathcal{S}_{2}$) as elementary resources. In fact, for
different proportions of these components one can see extermination or
stable coexistence (Tilman (1977)).

\[
\]

{\Large V Extinction of giant tortoises in Galapagos }

\[
\]

Concerning to evolution and selection, Galapagos islands are still a
prolific resource of study in ecology (Grant (2004)). For instance, native
giant tortoises $T$ \ live in Galapagos islands after centuries exploiting
its natural resources ($\mathcal{S}>T$). With the arrivals of humans being ($%
H$)\ some invaders ($I$), like goats and pigs, were introduced in the
islands and reaching a wild state. Invaders and tortoises, considered as
potential competitors (7),\ compete for primary resources ($\mathcal{S}%
>\left( T\supset \subset I\right) $). There is a real risque for extinction
of species $T$ in Galapagos. From 1995, international agencies are
developing an eradication program (Isabela project)\ to eliminate invaders $I
$ and re-introducing species $T$. In our symbolic notation: $\mathcal{S}%
>\left\{ \left( T\supset \subset I)\oplus (I\supset \subset H_{T}\right)
\right\} $. Where $H_{T}$ is the human being re-introducing species $T$ and
not compete with. Under this conditions, Gause's exclusion conjecture (8)
could be applied as a first approach and we have the four final process: (a) 
$\mathcal{S}>T$, (b) $\mathcal{S}>I$, (c) $\mathcal{S}>H_{T}$ (d) $\mathcal{S%
}>\left( H_{T}\oplus T\right) $. So, species $T$ is in three (of four)
possible final scenarios. Note that if $H_{T}\Leftrightarrow \phi $ (no
eradication) species $T$ occupies only one of two final scenarios and then
with a mayor risque to extinction.

\[
\]

{\Large VI Apparent competition }

\[
\]

It is well known (Holt (1984), Jeffries \& Lawton (1985)) that sometimes
there is apparent competition between two kind of no-competitive species.
This is due to the intervention of another (ignored) species which is a real
competitor. Examples of this curious process could be found in Begon \textit{%
et al} (1996). Let $A$ and $B$ be no-competing species exploiting a resource 
$\mathcal{S},$ namely, $\mathcal{S}>\left( A\oplus B\right) $. Let $C$ be a
potential competitor (of both) introduced in the system, and consider the
process:

\begin{equation}
\mathcal{S}>\left\{ \left( A\supset \subset C\right) \oplus \left( C\supset
\subset B\right) \right\} .
\end{equation}
Like to section V, a direct application of the basic tool (8) permits to
obtain the four mutual excluding possibilities:

(a) $\mathcal{S}>C$.

(b) $\mathcal{S}>\left( A\oplus B\right) $.

(c) \ $\mathcal{S}>A.$

(d) \ $\mathcal{S}>B$. \ \ \ \ \ \ \ \ \ \ \ \ \ \ \ \ \ \ \ \ \ \ \ \ \ \ \
\ \ \ \ \ \ \ \ \ \ \ \ \ \ \ \ \ \ \ \ \ \ \ \ \ \ \ \ \ \ \ \ \ \ \ \ \ \
\ \ \ \ \ \ \ \ \ \ \ \ \ \ \ \ \ \ \ \ \ \ \ \ \ \ \ \ \ \ \ \ \ \ \ \ \ \
\ \ \ \ \ \ \ \ \ \ \ \ \ \ \ \ \ \ \ \ \ \ \ \ \ \ \ \ \ \ \ \ \ \ \ \ \ \
\ \ \ \ \ 

The process (a) and (b) correspond to the elimination of $\left( A\oplus
B\right) $ or $C$. The more interesting cases ((c) and (d)) could be
interpreted erroneously as the result of competition between species $A$ and 
$B$ when the existence of the true competitor $C$\ is ignored. The process
(10) will be written in an abbreviate way as:

\begin{equation}
(10)\Leftrightarrow \mathcal{S}>\left( A\supset \subset C\supset \subset
B\right) .
\end{equation}

For more general cases like $\mathcal{S}>\left\{ A\supset \subset B\supset
\subset C...\supset \subset Z\right\} ,$ which does not contain some direct
competitors like $A$ and $Z$, the result is the same. So, the existence of
(ignored) intermediate competitors could be interpreted erroneously as
apparent competition.

\[
\]

{\Large Conclusions:}

\[
\]

Gause's exclusion conjecture was put in a operative symbolic nomenclature
(III). It could be applied to different cases like more than one resource or
more than two species (IV). The case of apparent competitors was also
explicitly treated (VI). Other cases like exclusion with sterile
individuals, regional \ displacement and others, could also be considered
(Flores 2005). In resume, this theoretical framework (ECOSET) corresponds to
an axiomatic theory to describe, and predict, ecological process.

\[
\]

\textbf{Acknowledgments:} This work was completed in the frame of \textit{%
Math Modelling in Galapagos Project (UTA-CIHDE-USFQ)}. It was supported by
UTA-Mayor project 4722 (Mathematical Modelling). I\ acknowledge the kind
hospitality of the GAIAS station at Galapagos (San Cristobal Island).

\[
\]

{\Large Resume for symbols}

\[
\]

$\supset $ \ Perturbation (no depredation).

$>$ \ Depredation.

$\oplus $ \ Two independent species in a region (eventually independent
process).

$\otimes $ \ Two interdependent species.

$\Rightarrow $ Temporal evolution.

$\Leftrightarrow $ Equivalence.

$\supset \subset $ \ \ Abbreviation for competition.

$><$ \ Mutual depredation.

$or$ \ Exclusion (some times $\vee $).

$\sum $ \ Independent species (eventually process): $A\oplus B\oplus C\oplus
D\oplus ...$.

\ 

\[
\]

\end{document}